\documentclass[twocolumn]{aastex701}

\DeclareRobustCommand{\okina}{%
  \raisebox{\dimexpr\fontcharht\font`A-\height}{%
    \scalebox{0.8}{`}%
  }%
}

\begin{document}

\title{Discovery and Analysis of a Type\,II Supernova Candidate at {\it z} = 3.19 from {\it JWST}'s COSMOS-Web Survey} 

\author[orcid=0009-0001-7491-9904, gname='Valeria', sname='Aparicio Diaz']{Valeria Aparicio}
\affiliation{Institute for Astronomy, University of Hawai\okina i at Manoa, 2680 Woodlawn Dr., Honolulu, HI 96822, USA}
\affiliation{Department of Physics and Astronomy, University of Hawai\okina i at Manoa, 2505 Correa Road, Honolulu, HI 96822, USA}
\email[show]{vaparici@hawaii.edu}  

\author[0000-0002-6230-0151]{David~O.~Jones}
\affiliation{Institute for Astronomy, University of Hawai\okina i, 640 N. A'ohoku Pl., Hilo, HI 96720, USA}
\email[show]{dojones@hawaii.edu}

\author[0000-0003-3953-9532]{Willem~B.~Hoogendam}
\altaffiliation{NSF Graduate Research Fellow}
\affiliation{Institute for Astronomy, University of Hawai\okina i at Manoa, 2680 Woodlawn Dr., Honolulu, HI 96822, USA}
\email{willemh@hawaii.edu}

\author[0000-0003-1169-1954]{Takashi~J.~Moriya}\affiliation{National Astronomical Observatory of Japan, National Institutes of Natural Sciences, 2-21-1 Osawa, Mitaka, Tokyo 181-8588, Japan}\affiliation{Graduate Institute for Advanced Studies, SOKENDAI, 2-21-1 Osawa, Mitaka, Tokyo 181-8588, Japan}\affiliation{School of Physics and Astronomy, Monash University, Clayton, VIC 3800, Australia}\email{takashi.moriya@nao.ac.jp} 

\author[0000-0003-4263-2228]{David~A.~Coulter}\affiliation{Space Telescope Science Institute, Baltimore, MD 21218, USA}\email{dcoulter@stsci.edu}

\author[0000-0002-2361-7201]{Justin~D.~R.~Pierel}\affiliation{Space Telescope Science Institute, Baltimore, MD 21218, USA}\email{jpierel@stsci.edu}

\author[0000-0003-2445-3891]{Matthew~Siebert}\affiliation{Space Telescope Science Institute, Baltimore, MD 21218, USA}\email{msiebert@stsci.edu}

\author[0000-0001-9269-5046]{Bingjie~Wang}
\thanks{NHFP Hubble Fellow}
\affiliation{Department of Astrophysical Sciences, Princeton University, Princeton, NJ 08544, USA}
\email{bjwang@princeton.edu}

\author[0000-0003-3596-8794]{Hollis~B.~Akins}
\email{hollis.akins@gmail.com}
\altaffiliation{NSF Graduate Research Fellow}
\affiliation{The University of Texas at Austin, 2515 Speedway Blvd Stop C1400, Austin, TX 78712, USA}

\author[0000-0002-0930-6466]{Caitlin~M.~Casey}
\email{cmcasey@ucsb.edu}
\affiliation{Department of Physics, University of California, Santa Barbara, Santa Barbara, CA 93106, USA}
\affiliation{Cosmic Dawn Center (DAWN), Denmark}

\author[0000-0003-4761-2197]{Nicole~E.~Drakos}
\email{ndrakos@hawaii.edu}
\affiliation{Department of Physics and Astronomy, University of Hawaii, Hilo, 200 W Kawili St, Hilo, HI 96720, USA}

\author[0000-0002-9382-9832]{Andreas L. Faisst}
\email{afaisst@caltech.edu}
\affiliation{Caltech/IPAC, MS 314-6, 1200 E. California Blvd. Pasadena, CA 91125, USA}

\author[0000-0003-2238-1572]{Ori~D.~Fox}\affiliation{Space Telescope Science Institute, Baltimore, MD 21218, USA}\email{ofox@stsci.edu}

\author[0009-0006-3071-7143]{Aryana Haghjoo}
\email{aryana.haghjoo@email.ucr.edu}
\affiliation{Department of Physics and Astronomy, University of California, Riverside, 900 University Ave, Riverside, CA 92521, USA}

\author[0000-0002-3301-3321]{Michaela Hirschmann}
\affiliation{Institute of Physics, GalSpec, Ecole Polytechnique Federale de Lausanne, Observatoire de Sauverny, Chemin Pegasi 51, 1290 Versoix, Switzerland}
\affiliation{INAF, Astronomical Observatory of Trieste, Via Tiepolo 11, 34131 Trieste, Italy}
\email{michaela.hirschmann@epfl.ch}

\author[0000-0002-7303-4397]{Olivier~Ilbert}
\email{olivier.ilbert@lam.fr}
\affiliation{Aix Marseille Univ, CNRS, CNES, LAM, Marseille, France  }

\author[0000-0001-9187-3605]{Jeyhan~S.~Kartaltepe}
\email{jeyhan@astro.rit.edu}
\affiliation{Laboratory for Multiwavelength Astrophysics, School of Physics and Astronomy, Rochester Institute of Technology, 84 Lomb Memorial Drive, Rochester, NY 14623, USA}

\author[0000-0002-6610-2048]{Anton~M.~Koekemoer}
\email{koekemoer@stsci.edu}
\affiliation{Space Telescope Science Institute, Baltimore, MD 21218, USA}

\author[0000-0002-9489-7765]{Henry~Joy~McCracken}
\email{hjmcc@iap.fr}
\affiliation{Institut d’Astrophysique de Paris, UMR 7095, CNRS, and Sorbonne Université, 98 bis boulevard Arago, F-75014 Paris, France}

\author[0000-0001-5846-4404]{Bahram~Mobasher}
\email{bahram.mobasher@ucr.edu}
\affiliation{Department of Physics and Astronomy, University of California, Riverside, 900 University Ave, Riverside, CA 92521, USA}

\author[0000-0002-4410-5387]{Armin~Rest}
\email{arest@stsci.edu}
\affiliation{Space Telescope Science Institute, Baltimore, MD 21218, USA}\affiliation{Physics and Astronomy Department, Johns Hopkins University, Baltimore, MD 21218, USA}

\author[0000-0002-4485-8549]{Jason~Rhodes}
\email{jason.d.rhodes@jpl.nasa.gov}
\affiliation{Jet Propulsion Laboratory, California Institute of Technology, 4800 Oak Grove Drive, Pasadena, CA 91001, USA}

\author[0000-0002-4271-0364]{Brant~E.~Robertson}
\email{brant@ucsc.edu}
\affiliation{Department of Astronomy and Astrophysics, University of California, Santa Cruz, 1156 High Street, Santa Cruz, CA 95064, USA}
\email{brant@ucsc.edu}

\author[0000-0002-7087-0701]{Marko~Shuntov}
\email{marko.shuntov@nbi.ku.dk}
\affiliation{Cosmic Dawn Center (DAWN), Denmark} 
\affiliation{Niels Bohr Institute, University of Copenhagen, Jagtvej 128, DK-2200, Copenhagen, Denmark}
\affiliation{University of Geneva, 24 rue du Général-Dufour, 1211 Genève 4, Switzerland}

\newcommand{\snname}{SN~2023aeaf}
\begin{abstract}
The launch of the \emph{James Webb Space Telescope} ({\it JWST}) has enabled the discovery of a small but increasing sample of high-redshift core-collapse supernovae (CC\,SNe), which provide new tests of massive star evolution in the early Universe. In this study, we report the discovery of \snname\ in COSMOS-Web survey observations, which at $z = 3.195$ has one of the highest SN spectroscopic redshifts to date. 
Using two epochs of {\it JWST} photometry separated by $\sim$1~month in the rest frame, we photometrically classify SN~2023aeaf by comparing the {\it JWST} photometry to spectrophotometric CC~SN and Type Ia (SN\,Ia) models and UV observations of SNe from the {\it Swift} telescope, finding that \snname\ is highly likely to be a Type II SN. A spectrum of the SN$+$host galaxy was also obtained $\sim$30 rest-frame days after discovery but shows  no clearly identifiable SN features, with H$\alpha$ emission from the host potentially masking emission from the SN. 
Although the limited photometric coverage prevents strong constraints on the explosion properties, we find that the  data are most consistent with a $\sim12M_\odot$ progenitor with $\sim$0.5$M_{\odot}$ of circumstellar material.
We next use the host-galaxy spectrum and photometry to model the host spectral energy distribution (SED) using the {\tt Prospector} Bayesian inference framework.  We find that the host is a star-forming galaxy with a sSFR of $ \log_{10}(\rm sSFR/yr^{-1})= -10.17^{+0.13}_{-0.10}$, a stellar mass of $\log(M_\star/M_\odot) = 9.04^{+0.03}_{-0.04}$, and a gas-phase metallicity of  
$12 +{\rm log_{10}}({\rm O/H}) = 7.82\pm0.02$.   \snname\ joins a growing sample of early Universe CC~SNe with high luminosities, dense CSM, and low-metallicity environments.

\end{abstract}


\section{Introduction} 

Supernovae (SNe) trace the stellar populations and evolutionary histories of their host galaxies \citep[][]{Galbany14,Anderson15,Schulze_2021,Taggart21}. Core-collapse (CC) SNe in particular, as the endpoints of $M_\star  \gtrsim 8 M_\odot$ stars with lifetimes of only a few million years (see \citealp{Smartt_2009} for a review), trace recent star formation and the dynamical feedback that shapes that star formation \citep{James_2006}. The chemical yields from these explosions subsequently enrich and shape the surrounding interstellar medium, and influence the metallicity gradients within galaxies \citep{Baade&Zwicky, oppenheimer1939gravitational, Smartt_2009,Kong26}. The rates of CC\,SNe also provide an independent measurement of the cosmic star-formation rate density, and probe variations in the high-mass end of the stellar initial mass function (IMF; \citealp{Dahlen2012,Strolger15}). A better understanding of these explosions is therefore essential for linking their physical role in galaxy evolution, star formation, and metal enrichment to the signatures we detect in transient surveys. 

Among the diverse population of SNe, Type II SNe (SNe\,II) are the most common type, and are classified by the presence of strong hydrogen lines in their spectra due to a hydrogen-rich envelopes in their progenitor stars \citep{Morgan, fang2024diversityhydrogenrichenvelopemass}. SNe\,II exhibit a wide range of diversity in  observed properties, including in their light curves, spectra, and peak luminosities. This diversity arises from a range of physical factors, including the initial masses, metallicities, and rotation rates of their progenitors, as well as differences in their circumstellar environments \citep{heger2003massive,Valenti16,Arcavi17, martinez2022snIIdiversity}. 

The structure and evolution of the progenitor, particularly the extent of hydrogen-rich envelope retention, are strongly influenced by its metallicity. The retained envelope determines how the expanding shock wave interacts with the circumstellar material (CSM) and influences the duration of the cooling phase, the efficiency of energy deposition from radioactive decay, and the overall luminosity evolution. The effects of metallicity on mass-loss rates and final envelope masses therefore contribute to variations in light-curve shape and energy \citep{ Vink2001, Heger2003, dessart2013quantitative, smith2014}.

As low‑metallicity environments become increasingly common at higher redshift \citep{Larson1998, Sanders_2021}, this chemical evolution should directly shape the observable properties of CC\,SNe.  While progenitors' low metallicities are likely to systematically affect properties of the SN explosion \citep{Anderson18,Gutierrez20,moriya2025,Tucker2024_2023ufx}, the limited sample of observed high-redshift CC~SNe means that many of the predictions for early Universe CC\,SNe are untested. Expanding the census of such SNe  will be essential to understand the extent to which changes in progenitor structure in the early Universe systematically affect explosion energies, spectra, and photometric behavior across cosmic time.

Ground-based surveys have successfully identified several high-$z$ CC \, SNe, including superluminous SNe (SLSNe) and SNe IIn, at $z > 2$ \citep[e.g.,][]{Cooke09,Cooke12,Smith18,Curtin19,Moriya19,Gal-Yam19,Johansson25}, while {\it Hubble Space Telescope} surveys such as CANDELS and CLASH \citep{2011Grogin,Postman12} discovered CC \, SNe to $z \simeq 2.5$ \citep{Strolger15}.
However, with its IR imaging and spectroscopic capabilities, {\it JWST} has already discovered larger samples of distant SNe than was possible from all previous surveys combined \citep{decoursey2025jadestransientsurveydiscovery, yan25, fox2026}, and characterized in detail a handful of events with follow-up spectroscopy \citep[e.g.,][]{ Pierel24,siebert2024,coulter2025discovery,siebert25, Pierel25, coulter2026spectroscopicallyconfirmedstronglylensed}.  The sensitivity and near- to mid-IR coverage of {\it JWST} in particular enables multiwavelength, high-signal-to-noise (S/N) characterization of such SNe over longer time baselines than was previously possible.

Multiple deep {\it JWST} imaging surveys, including JADES  \citep{eisenstein2023JADES}, PEARLS \citep{Windhorst_2022pearls}, and COSMOS-Web \citep{casey2023cosmosweboverviewjwstcosmic}, discovered high-$z$ SNe in their survey fields. 
Of these, the largest-area survey is COSMOS-Web \citep{casey2023cosmosweboverviewjwstcosmic, fox2026}, which aims to map the universe’s large-scale structure, investigate galaxy evolution, and advance our understanding of reionization in the early universe.

Here, we report and analyze a likely SN\,II found in the COSMOS-Web survey located at R.A. = 10:00:22.8, decl.\ = +02:22:24.6.  The SN was discovered at AB magnitudes near $\sim27.5$ mag in the {\it F115W}, {\it F150W}, {\it F277W}, and {\it F444W} bands.  \snname\ had an initial photometric redshift of 3.55 and was initially found to be marginally consistent with a SN\,Ia at this redshift ($\chi^2/\nu = 3.85$) as reported by \citet{fox2026}.
Subsequent spectroscopic observations with NIRSpec measured its redshift to be $z = 3.195$;  at the time of writing, \snname\ has the fourth-highest spectroscopic redshift of any SN with more than one observational epoch (excluding template epochs), exceeded only by a small number of other {\it JWST}-discovered SNe \citep{decoursey2025jadestransientsurveydiscovery,coulter2025discovery, coulter2026spectroscopicallyconfirmedstronglylensed, fox2026}.

We organize this paper as follows.  In Section \ref{Data}, we summarize the observational details. Section \ref{classification_spectral_analysis} describes our methods for the photometric classification and spectroscopic analysis of \snname. Section \ref{host_galaxy} derives the host-galaxy properties by analyzing the spectral lines and fitting the Spectral Energy Distribution (SED). Finally, in Section \ref{sec:progenitor_modeling} we model the light curve to obtain progenitor properties. In Section \ref{discussions and conclusions} we conclude.

\section{Data} \label{Data}
\snname\ was discovered by the COSMOS-Web survey (PID 1727; PI: C. M. Casey, J. S. Kartaltepe) with details of the SN discovery program given in \citet{fox2026}.  COSMOS-Web is a 255-hour treasury survey covering 0.54 deg\textsuperscript{2} with {\it JWST}'s Near-Infrared Camera (NIRCam) over four filters (\textit{F115W, F150W, F277W, and F444W}), and covering 0.2~deg\textsuperscript{2} with MIRI.  COSMOS-Web achieved 5$\sigma$ point-source depths of approximately 27.5–28.2~mag \citep{casey2023cosmosweboverviewjwstcosmic, fox2026},  with its first observations occurring during Cycle 1 between December 2022 and January 2023; these data provided an initial epoch for transient searches. A second epoch was obtained later in 2023, which enabled the identification of numerous transient candidates. Among these, \snname\ was identified as a candidate high-redshift SN due to its host-galaxy photometric redshift of $z = 3.55$ and magnitude consistent with a high-$z$ SN. Its discovery then motivated a {\it JWST} Director’s Discretionary Time program (6585; PI: D. Coulter), which enabled follow-up imaging with NIRCam and spectroscopy with NIRSpec. The full set of data is described below.  

\begin{figure*}
    \centering
    \includegraphics[width=0.8\linewidth]{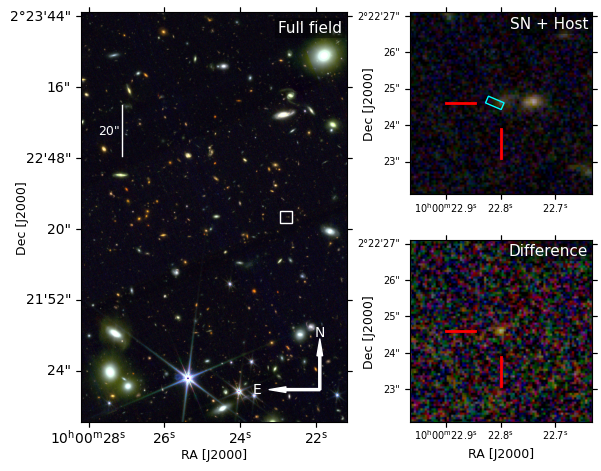}
    \caption{Left: Full-field {\it JWST}/NIRCam RGB image constructed using the {\it F115W} (blue), {\it F277W} (green), and {\it F444W} (red) filters. The white box marks the location of \snname. Top Right: Zoomed-in RGB image of the SN and host galaxy corresponding to the boxed region in the left panel. The cyan rectangle shows the orientation and position of the {\it JWST}/NIRSpec MSA slit used for the prism spectroscopy. Bottom Right: Difference image created by subtracting template images from the SN$+$host image shown in the top-right panel, highlighting the transient emission from \snname.}
    \label{fig:jwst-image}
\end{figure*}

\subsection{Photometry} \label{photometry}
The photometric data consist of 2 epochs of NIRCam imaging obtained in the \textit{F115W, F150W, F277W}, and {\it F444W} filters, which are separated by an average of 126 days (30~days in the rest frame of \snname). We perform photometry following the procedures developed in \citet{siebert2024,coulter2025discovery,decoursey2025jadestransientsurveydiscovery}. To summarize, individual Level 2 (CAL) exposures are aligned with JHAT \citep{2023Rest} and drizzled into Level 3 mosaics with the {\it JWST} pipeline (version 1.13.4). Difference images are then generated by using the High Order Transform of PSF and Template Subtraction ({\tt HOTPANTS}; \citealp{Becker2015_HOTPANTS})\footnote{\url{https://github.com/acbecker/hotpants}.} software, which convolves the template image to match the point spread function (PSF) of the survey image and subtracts it to isolate the transient flux. The photometry was extracted with the {\tt space\_phot}\footnote{\url{space-phot.readthedocs.io}.} PSF-fitting routine on 5×5 pixel cutouts centered at the SN position. The PSF models were derived from spatially and temporally dependent {\tt webbpsf}\footnote{\url{ https://webbpsf.readthedocs.io}.} simulations and drizzled to match the mosaics. Figure \ref{fig:jwst-image} shows three-color images of the  discovery epoch for \snname, while Table~\ref{tab:photdata} presents the {\it JWST} photometric measurements of \snname\, including observation dates, filters, and apparent magnitudes with uncertainties.

\begin{table}
\centering
\caption{Summary of NIRCam photometric measurements.}

\begin{tabular}{ccc}
\hline
MJD & Filter & $m_{AB}$ \\
&&(mag)\\
\hline
60305.3594 & F115W & $27.56 \pm 0.20$ \\
60303.9660 & F150W & $27.51 \pm 0.13$ \\
60302.6023 & F277W & $27.77 \pm 0.15$ \\
60301.6635 & F444W & $27.62 \pm 0.20$ \\
\hline
60429.2563 & F115W & $29.21 \pm 0.34$ \\
60429.2761 & F150W & $30.35 \pm 0.60$ \\
60429.2763 & F277W & $29.45 \pm 0.33$ \\
60429.2563 & F444W & $29.19 \pm 0.39$ \\
\hline
\end{tabular}

\label{tab:photdata}
\end{table}

\begin{figure}
    \centering
    \includegraphics[width=1.0\linewidth]{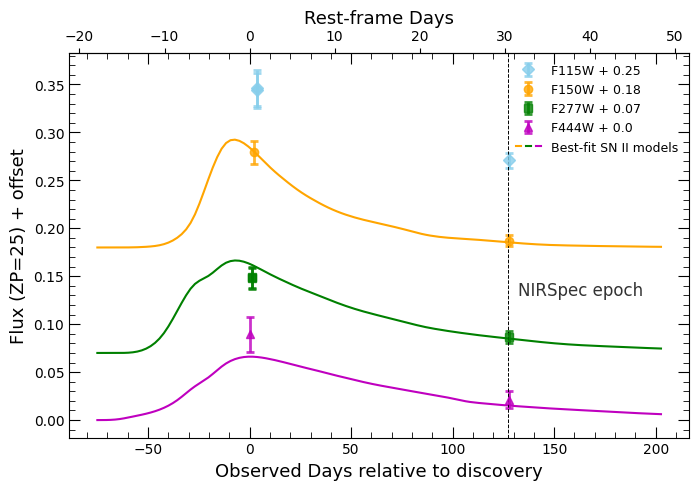}
    \caption{Observed photometry of \snname.
    Solid lines correspond to the best-fit SN\,II template. The fit excludes the {\it F115W} band, which is too blue to be fit by our templates, and thus no solid line is displayed for this band (see Section \ref{classification}).}
    \label{fig:f115w_ref}
\end{figure}

\subsection{Spectroscopy} \label{spectroscopy}
%crazy how much info you can find in fits headers lol 
We obtained a spectrum of \snname\ and its host galaxy on MJD $60429$ using NIRSpec in Multi-Object Spectroscopy (MOS) mode. 
The data were collected on April 29, 2024 as part of Director’s Discretionary Program 6585 (PI: D. Coulter).
\snname\ was observed with the PRISM/CLEAR configuration, providing continuous wavelength coverage from 0.6 to 5.3 $\mu m$ with a single micro-shutter ($0.46'' \times 0.20''$), and the exposure was performed in a 3-shutter nod pattern to enable accurate background subtraction. NIRSpec was read out in the NRSIRS2 mode. The final combined product includes 9 exposures, corresponding to an effective on-source integration time of 17725.5 s ($\sim$4.9 hr).

We processed the data using the {\it JWST} Science Calibration Pipeline (version 1.13.4) with CRDS context {\tt jwst\_1242.pmap}. To assess the reliability of the extracted spectrum, we inspected the two-dimensional (s2d) products to verify that the spectral trace was free from contamination by nearby sources or detector artifacts, and that the background subtraction was performed correctly. These checks ensure that the resulting one-dimensional spectrum accurately represents the target source.

A summary of the spectroscopic observation details is given in Table \ref{tab:spectra} and the spectrum is shown in Section \ref{sec:sn_spectra}.

\begin{table}
\centering
\caption{NIRSpec observing setup.}
\begin{tabular}{ll}
\hline\hline
Instrument & NIRSpec \\
Mode & MOS (MSA Spectroscopy) \\
Wavelength Range & 0.6--5.3 $\mu$m \\
Grating/Filter & PRISM/CLEAR \\
Spectral Resolution & $R \sim 30$--300 \\
Readout Pattern & NRSIRS2 \\
Exposures Included & 9 \\
Nods & 3 (3-shutter-slitless pattern) \\
Per-Exposure Integration & 656.5 s \\
Total Exposure Time & 17725.5 s (4.9 hr) \\
Elapsed Duration & 18119.4 s \\
Detector/Subarray & NRS1, FULL \\
\hline
\label{tab:spectra}
\end{tabular}
\end{table}

\section{Classification and Spectral Analysis}\label{classification_spectral_analysis}

\subsection{Photometric Classification} \label{classification}

We use the photometric data listed in Table~\ref{tab:photdata} and shown in Figure~\ref{fig:f115w_ref} to determine the type of \snname\ using {\tt STARDUST2} \citep{rodney_stardust2}. {\tt STARDUST2} is a Bayesian classifier originally designed for high-redshift SNe discovered with the {\it Hubble Space Telescope} ({\it HST}). 
Its method classifies a given SN by comparing its light curve to simulated populations of SNe. These simulations are generated from spectrophotometric templates, including 27 Type II and 15 Type Ib/c templates compiled in the SuperNova ANAlysis package \citep{Kessler09}, together with the NIR-extended SALT3 model for SNe Ia \citep{Pierel_2018_class,Pierel_2022}.  Realizations of the templates are simulated using luminosity functions from \citet{Li2011LOSS_LF}, dust distributions as outlined in \citet{Rodney14_rates}, and skewed normal distributions for the SN\,Ia shape and color parameters. Due to the paucity of high-cadence, high-S/N SN observations at high redshift, we note that all templates with the exception of SALT3-NIR are currently constructed from $z \lesssim 0.2$~SNe; although {\tt STARDUST2} has proven effective at high-$z$ SN classifications \citep[e.g.,][]{decoursey2025jadestransientsurveydiscovery}, the lack of high-$z$ template representation is an unavoidable caveat to the analysis below.  

{\tt STARDUST2} compares these simulations to the data, and marginalizes over all simulation templates/parameters simultaneously to determine the posterior probability that a given SN is Type Ia, Ib/c, or II. We allow the time of maximum light ($t_{\rm max}$) to vary between 50 rest-frame days before the first photometric epoch and 50 days after the final epoch. This broad prior is a conservative limit given the data: the SN was faintest (or pre-explosion) during the template epoch, which was taken $\sim$50 rest-frame days prior to discovery, and the light curve declined within the $\sim$30 rest-frame days between the discovery epoch and the final epoch.
 
We also restrict the input data for \snname\ to rest-frame wavelengths $>3000$~\AA, due to poor reliability of the CC\,SN templates blueward of this wavelength.  The CC\,SN templates were constructed by mangling spectroscopic time series to match optical photometric data;  however, most photometric and spectroscopic data did not extend bluer than the $U$ or $u$-band ($\lambda_{eff} \simeq 3600-3700 {\rm \AA}$). Therefore, before fitting the light-curve, we omitted the bluest filter ($F115W\approx2700$ \AA\ in the rest frame) from the light‑curve fit. We revisit these UV data by comparing to {\it Swift}-observed SNe in Section \ref{sec:swift}.

\begin{figure*}
    \centering
    \includegraphics[width=1.0\linewidth]{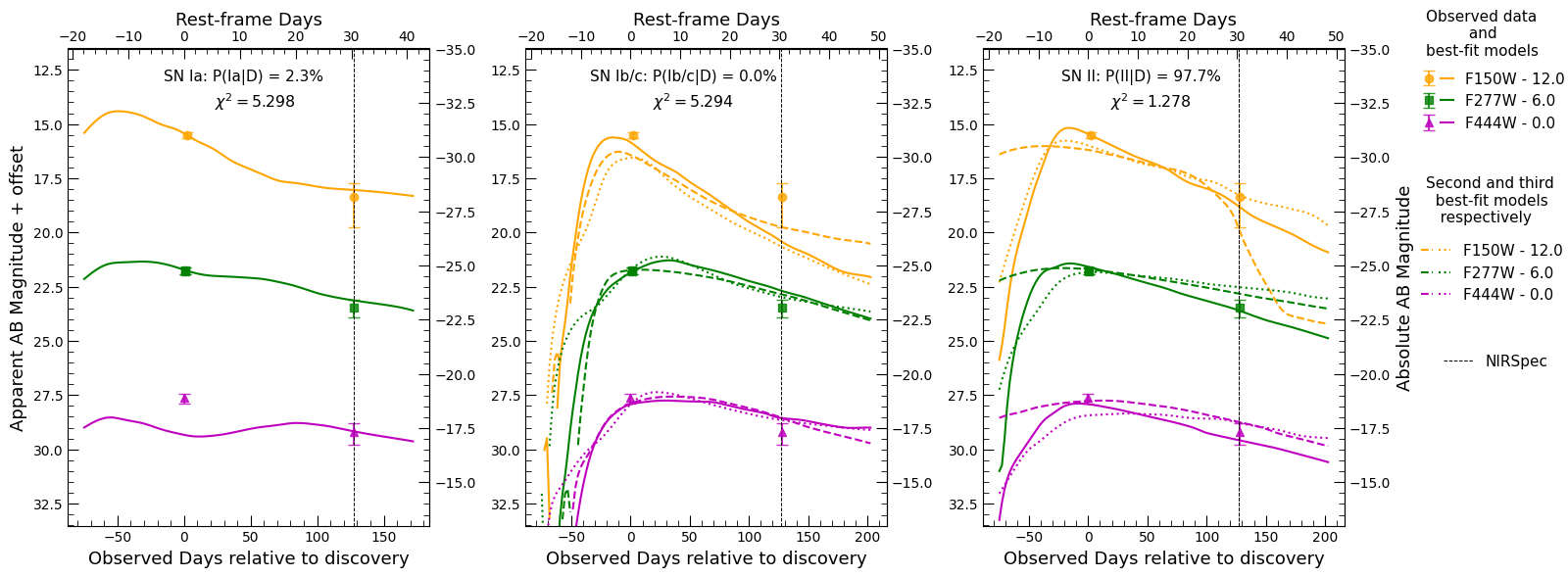}
    \caption{Left: The multiband light curve of \snname\ (colored points, with offsets applied for clarity) along with the best-fit SN Ia SALT3-NIR model shown as continuous curves.  The epoch of the NIRSpec observation is indicated by the vertical dotted black line.  Middle: Similar to the left panel but highlighting the best-fit SN\,Ib/c model. For comparison, the second- and third-ranked fits are overplotted as dashed and dotted curves, respectively. Right: Similar to the middle panel but showing the best-fit SN\,II models.  Labels show the reduced $\chi^2$ of the best-fit template for each type, and posterior probabilities that marginalize over all templates.} 
    \label{fig:classification_lightcurve}
\end{figure*}

{\tt STARDUST2} shows that \snname is most consistent with a SN\,II, with the SN~2007pg template having the highest individual probability. The Bayesian classification yields a probability of 97.2\% for the SN\,II scenario, while the other probabilities are small $(<3\%)$.  For the best-fit SN\,II model, $\chi^2/\nu=1.239$, while for the best-fit SN\,Ia and SN\,Ib/c models $\chi^2/\nu = 5.267$ and $\chi^2/\nu=5.235$, respectively.
%qualitatively 
Figure~\ref{fig:classification_lightcurve} compares the best-fit models for all three SN types, along with the second- and third-ranked fits for Types Ib/c and II. 

From Figure \ref{fig:classification_lightcurve}, we see from the SN~Ia model that the light curve lacks the secondary maximum in the NIR typical of a SN\,Ia, while the SALT3 color parameter is unusually blue ($c = -0.23$); this color is within standard SN\,Ia cosmology cuts but bluer than all but one of the $\sim$1700 SNe\,Ia in the Pantheon$+$ compilation  \citep{Scolnic22}. We can observe a reasonable match with some SNe Ib/c templates; however, their colors are generally too red, or their decline rates are too fast in the {\it F150W} band. 
SN II models are a consistent match to the observed light-curves across filters, reproducing the evolution and fluxes successfully. While it is possible that {\tt STARDUST2} overestimates the SN\,II probability due to a limited Ib/c template set --- visually, the best-fit SN\,Ibc template appears to provide a reasonable match, while the other template fits are more discrepant --- we find that the data align most closely with the SN~II scenario. 

Using the best-fit SN\,II model, we estimate a peak absolute magnitude of $M \sim -18.95$ in the rest-frame $r$ band.
Comparing to the $R$-band CC\,SN luminosity functions from the Lick Observatory Supernova Search (LOSS; \citealp{Li2011LOSS_LF}), \snname\ is $\sim$2.8$\sigma$ above the mean luminosity for Type II-P SNe, and $\sim$2.5$\sigma$ above the mean II-L luminosity.  It is consistent with SN\,IIn luminosities, although its decline rate is faster than typical for this subtype \citep[e.g.,][]{Nyholm20}. However, because \snname\ was discovered in a magnitude-limited survey, we would expect systematically brighter discoveries compared to a volume-limited survey; $\sim$5\% of SNe\,II  in the ASAS-SN CC~SN sample have comparable or brighter absolute magnitudes than \snname\ \citep[their Figure 9]{Pessi25}.  However, it is also possible that SNe\,II have higher luminosities in low-metallicity environments, which is consistent with the low-metallicity SNe~2023ufx \citep{Ravi2025_2023ufx, Tucker2024_2023ufx}, 2023adsv \citep{coulter2025discovery}, and the six high-$z$ SNe\,II analyzed by \cite{moriya2025}; all report a similar trend of elevated luminosities in low-metallicity hosts.

\subsection{Swift SN sample comparison}
\label{sec:swift}

The {\it Neil Gehrels Swift Observatory} is a multiwavelength space telescope originally designed to study gamma-ray bursts, but it has since become one of the most productive space telescopes for SN follow-up observations \citep{Gehrels2004_SWIFT}. Using data from the \textit{Swift Ultraviolet/Optical Telescope} ({\it UVOT}), we compare the UV–optical color evolution of \snname\ to the {\it UVW1}$– u$ colors of different SN subtypes over time. The Swift SOUSA \citep{Brown_2014} archive contains hundreds of well-sampled nearby SNe, which we use to construct mean UV color curves for each SN type.

\begin{figure*}
    \centering
    \includegraphics[width=\linewidth]{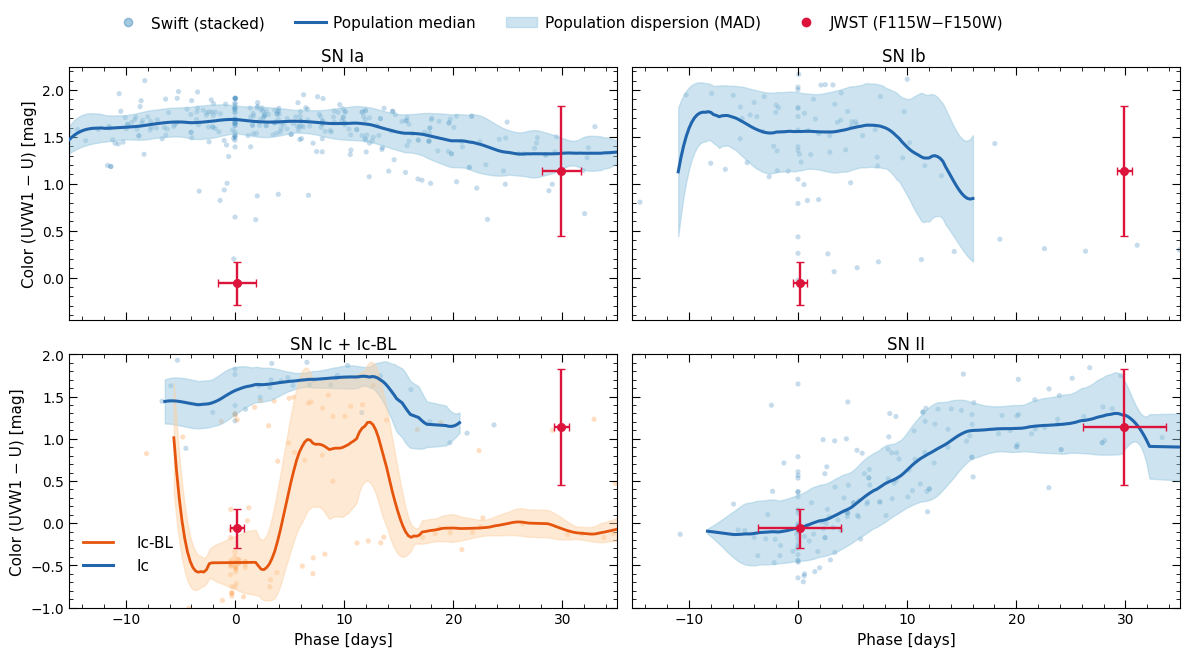}
    \caption{Comparison between Swift UV–optical color evolution of different SN subtypes and our {\it JWST} observations.
Each panel shows the population-averaged Swift color curves (solid blue lines) and $1\sigma$ dispersions (shaded regions) for different SN subtypes (Type Ia, Ib, Ic, and II). The y axis shows the Swift {\it UVW1}$– u$ color, corresponding to central wavelengths of 2600 \AA\ and 3465 \AA\ respectively, as a function of time since optical maximum light. Red data points show the {\it JWST} color ({\it F115W}$-${\it F150W}) and are plotted with respect to the discovery time. Horizontal error bars reflect the uncertainty in the fitted time of maximum light derived from the corresponding best-fit {\tt STARDUST2} template. At $z = 3.19$, these bands correspond to central rest-frame wavelengths of approximately 2750 \AA\ and 3580 \AA, respectively.}
    \label{fig:swift_comp}
\end{figure*}

In Figure \ref{fig:swift_comp}, we show that our data are most consistent with the SN II population by comparing the {\it JWST} {\it F115W}$-${\it F150W} color, which probes rest-frame wavelengths of approximately 2750 and 3580~\AA, to the Swift {\it UVW1}$-u$ color, which samples similar rest-frame wavelengths of approximately 2600 and 3465~\AA. Because the inferred epoch of maximum light depends on the adopted SN template, the {\it JWST} color point may shift along the phase axis within the uncertainties of the fitted time of maximum. However, even allowing for these phase uncertainties, the measurements remain largely inconsistent with the UV colors of other SN types.

These results provide additional support for the photometric classification shown in Figure \ref{fig:classification_lightcurve}. There is no overlap with SN\,Ia UV colors at maximum light, and even the bluest SN\,Ia subtypes are inconsistent with this object at maximum light in the $UVW1-u$ color \citep[see also][]{Hoogendam_2024}. Although there is some overlap with SN Ib/c models, only one SN Ic in the Swift comparison sample provides a similarly blue color near maximum light, whereas the majority of SNe~II align well with the observations. 

In summary, both the {\tt STARDUST2} classifications and the UV colors favor the SN\,II scenario for \snname, although it is difficult to fully rule out the possibility of a SN\,Ib or Ic. However, we note that this comparison is necessarily limited by the assumption that high-redshift SNe\,II exhibit color evolution similar to their low-redshift counterparts, as both the Swift comparison sample and the classification templates in Section \ref{classification} are derived from low-$z$ SNe.

\subsection{Spectral Analysis}\label{sec:sn_spectra}

We next attempt to isolate the SN spectrum by subtracting from the SN$+$host galaxy spectrum a model spectrum of the host determined using spectral energy distribution (SED) fitting of the host galaxy photometry. We use host-galaxy photometry from the COSMOS collaboration \citep{Shuntov25}, which includes {\it JWST} imaging \citep{casey2023cosmosweboverviewjwstcosmic} in the {\it F090W}, {\it F115W}, {\it F150W}, {\it F200W}, {\it F277W}, {\it F356W}, {\it F410M}, and {\it F444W} filters, Subaru Hyper Suprime-Cam imaging \citep{AIHARA2022} in {\it grizy}, and {\it HST} imaging \citep{2007koekemoer, 2011koekemoer, 2011Grogin} in the {\it F606W}, {\it F814W}, {\it F125W}, and {\it F160W} filters (see Figure 3 in \citealp{Shuntov25} for details of the photometric measurements). 

We used {\tt Prospector} \citep{leja_prospector,johnson_prospector} to fit these data with stellar population synthesis models \citep{conroy_2009,conroy_2010}.  {\tt Prospector} uses a Bayesian methodology that allows us to constrain galaxy properties by modeling the high-dimensional, correlated parameter space of galaxy SEDs by marginalizing over poorly constrained parameters, choosing well-informed priors, and robustly propagating uncertainties.  Prospector enables a joint determination of stellar populations, star formation history, dust attenuation, AGN contributions, and nebular emission. We use the {\tt Prospector-$\alpha$} model, which adopts a binned nonparametric star-formation history.  See \citet[their Section 2.2.5 and Appendix A]{dojo24} for additional details, as well as the complete list of parameters and priors used in this analysis.
 
\begin{figure}
    \centering
    \includegraphics[width=1.0\linewidth]{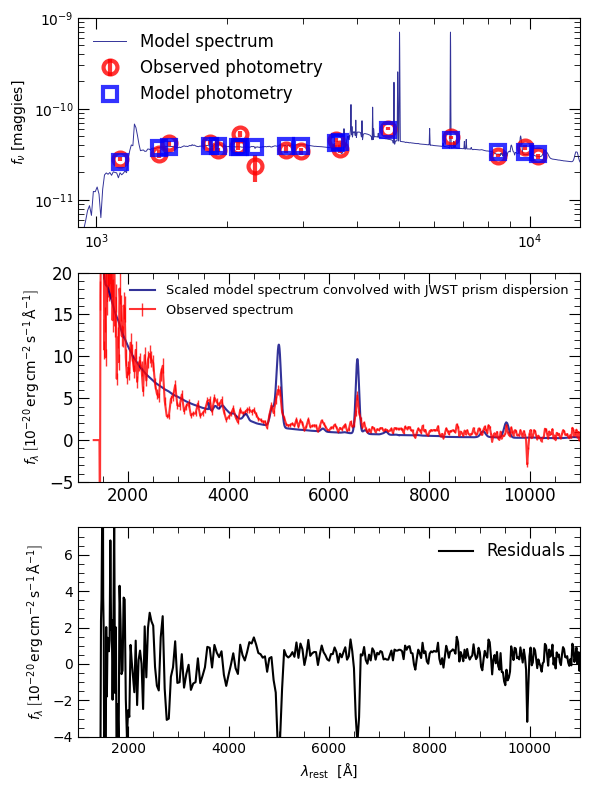}
    \caption{Comparison between the {\tt Prospector} SED model, derived using photometric data only, and the observed photometric and spectroscopic data. Top: Best-fit model spectrum (blue line) plotted in the rest frame, along with observed photometry (red circles) and model-predicted photometry (blue squares). Middle: {\it JWST}/NIRSpec prism spectrum (red) compared to the model spectrum after convolving it with the wavelength-dependent NIRSpec line-spread function (blue). Bottom: Residuals computed by subtracting the convolved  model spectrum from the observed spectrum.  All wavelengths are shown in the rest frame.}
    \label{fig:spectra}
\end{figure}

We convolve the {\tt Prospector}-derived best-fit model with the {\it JWST} NIRSpec prism dispersion and resample it onto the observed wavelength grid before comparing it to our observed SN$+$host spectrum. 
We scale the model spectrum to match the data using a least-squares fit. 
We then subtract the scaled host-galaxy spectral model from the SN$+$host spectrum to isolate the spectrum of the SN. The results of this analysis are shown in Figure~\ref{fig:spectra}.

The residual spectrum shown in the bottom panel of Figure \ref{fig:spectra} does not appear to contain significant SN continuum flux or spectral features. 
Although we do not detect clear evidence of a broad $H\alpha$ feature or other signatures of SN emission in the spectrum, given the relative magnitude of the galaxy versus the SN within the NIRSpec slit (see Section \ref{host_galaxy} below), it is likely that such features would be too faint to detect in our data.
While the model slightly over-predicts the strength of narrow emission lines, this is not unexpected, as the photometric fit represents the integrated properties of the host galaxy, whereas the NIRSpec slit samples a localized region that may not be fully representative.

\section{Host-Galaxy Analysis} \label{host_galaxy}

To characterize the physical properties of the host galaxy, we first analyze emission lines in the NIRSpec spectrum to constrain the dust extinction and metallicity. Second, we perform spectral energy distribution (SED) fitting using {\tt Prospector}, adopting the same model and priors as in the previous analysis, but now jointly fitting both the photometry and the {\it JWST} prism spectrum to better constrain galaxy properties including the stellar mass, star formation rate, metallicity, and dust content.

As an empirical estimate of the potential SN contamination in this spectrum, we compare the SN $F277W$ (rest-frame $\sim$6600\AA) magnitude coincident with the spectral epoch of $29.45\pm0.33$~mag to the host-galaxy magnitude in the same filter measured within a 0.2\arcsec\ aperture at the SN location, which we measure to be $\sim26.6$~mag. This implies that the host galaxy is brighter by a factor of $\sim$14 even in the filter encompassing H$\alpha$.
Based on this estimate, and the lack of broad emission features in the residual spectrum after host-model subtraction (Figure \ref{fig:spectra}), we assume that line contamination from the SN does not significantly bias our results.

\subsection{Spectral Line Analysis} \label{host_spectral_analysis}
We first estimate the dust attenuation and gas-phase metallicity of the host galaxy using emission-line diagnostics. To do so, we fit prominent nebular emission lines in the JWST/NIRSpec spectrum, accounting for line blending at the instrument resolution. Prior to fitting the emission lines, we subtract a local continuum from the spectrum to remove the underlying host galaxy stellar continuum and any smooth continuum contribution from the SN. For blended features, such as H$\beta$ and the [O\,III] $\lambda\lambda4959,5007$ doublet, we model H$\beta$ with a single Gaussian profile, while the [O\,III] doublet is simultaneously fit with an additional Gaussian (effectively treating it as a single broadened feature at the NIRSpec resolution). For the H$\beta$+[O\,III] fit, the amplitudes, centroids, and widths of both Gaussian components were allowed to vary freely, with loose priors placed on the expected observer-frame central wavelengths of the features. The line fitting is performed using the {\tt emcee} Python package \citep{Foreman_Mackey_2013}, an affine-invariant Markov Chain Monte Carlo (MCMC) sampler.

With these line measurements, we employ spectral diagnostics to constrain the metallicity and dust extinction. First, we compute the $R_{23}$ line ratio, defined as

\begin{equation}
R_{23} = \frac{[{\rm O\,II}]\lambda3727 + [{\rm O\,III}]\lambda\lambda4959,5007}{H\beta},
\end{equation}

\noindent using the measured emission-line fluxes. We note that $[{\rm O\,II}]\lambda3727$ is a somewhat marginal detection with a ${\rm S/N} \simeq 4.5$ while $H\beta$ and $[OIII]$ are more confidently detected. 
We find $R_{23} = 7.22^{+3.01}_{-1.64}$.

We convert the measured $R_{23}$ value to an oxygen abundance using the calibration of \citet{Mcgaugh1991}, adopting the analytic formulation presented by \citet{kobulnicky1999}. This yields $12+\log(\mathrm{O/H}) = 8.07^{+0.29}_{-0.20}$.  However, because this calibration is based on
local HII regions, we also estimate the abundance using the redshift-dependent
calibration of \cite{hirschmann2023}. This gives
$12+\log(\mathrm{O/H}) = 7.535^{+0.293}_{-0.203}$ (as the relation is bimodal, we adopt the lower-metallicity branch; this gives better consistency with the local $R_{23}$ calibration and the {\tt Prospector} results).

Second, we estimate the host-galaxy dust extinction using the Balmer decrement to inform subsequent modeling of the SN in Section~\ref{sec:progenitor_modeling}. The observed H$\alpha$/H$\beta$ ratio is measured to be $2.86^{+1.21}_{-0.64}$, consistent within uncertainties with the theoretical case~B recombination value of 2.86 assuming no dust extinction. This corresponds to an extinction value of $E(B-V) = 0.02^{+0.33}_{-0.24}$~mag, indicating no significant dust attenuation, though the uncertainty remains relatively large.

\subsection{Spectral Energy Distribution Fitting with Prospector}

\begin{figure}
    \centering
    \includegraphics[width=0.9\linewidth]{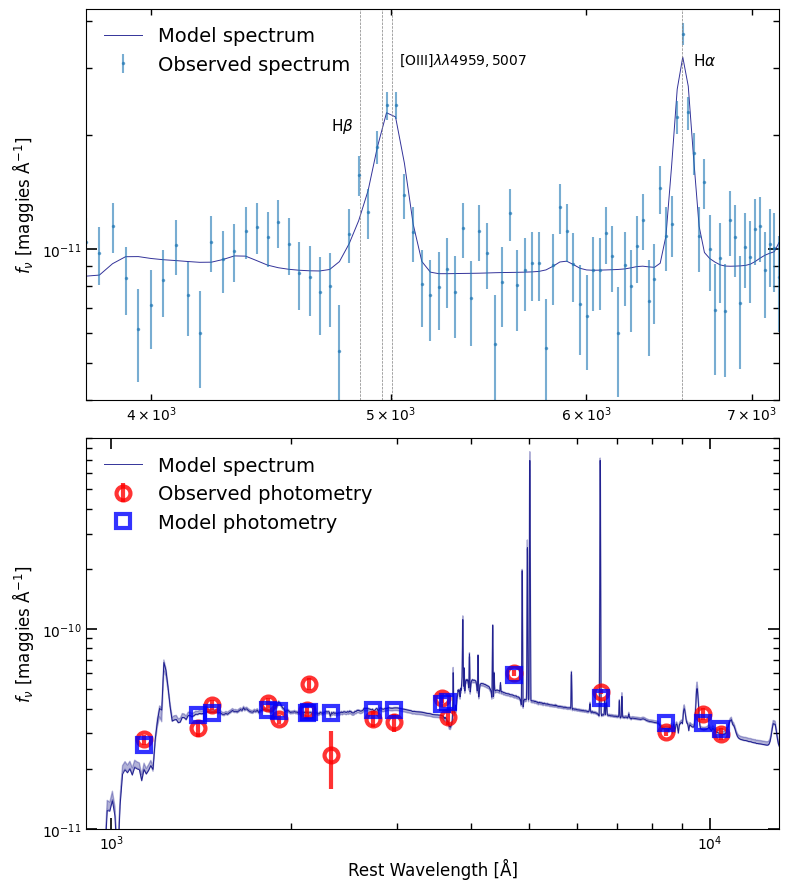}
    \caption{Top panel: Comparison between the observed host-galaxy spectrum (blue points) and the best-fit model spectrum from SED fitting after including the spectrum in the {\tt Prospector} fit (solid line). 
Bottom panel: Best-fit host-galaxy SED model (blue solid line), determined using both host-galaxy photometry and spectra. Red circles show the observed  photometry, while blue squares indicate the corresponding model photometry derived from the best-fit SED.  Light shading indicates errors on the best-fit spectral model.}
    \label{fig:spec_phot_comp}
\end{figure}

We again use {\tt Prospector} to fit the SED, but include the {\it JWST} prism spectrum in the parameter estimation.  We allow a flux offset as an additional parameter, to account for the difference between the host-galaxy photometric apertures and the prism extraction, and include the {\it JWST} prism dispersion function so that the model SED can be convolved to match the data during parameter estimation.  Here, the nebular emission lines in the spectrum are fit with a grid of Cloudy models \citep{osterbrock2006} that use Fast Stellar Population Synthesis (FSPS) stellar populations as the source of ionizing radiation \citep[see][]{Byler17}.  The resulting best-fit SED is shown in Figure \ref{fig:spec_phot_comp}.

\begin{figure*}
    \centering
    \includegraphics[width=0.9\linewidth]{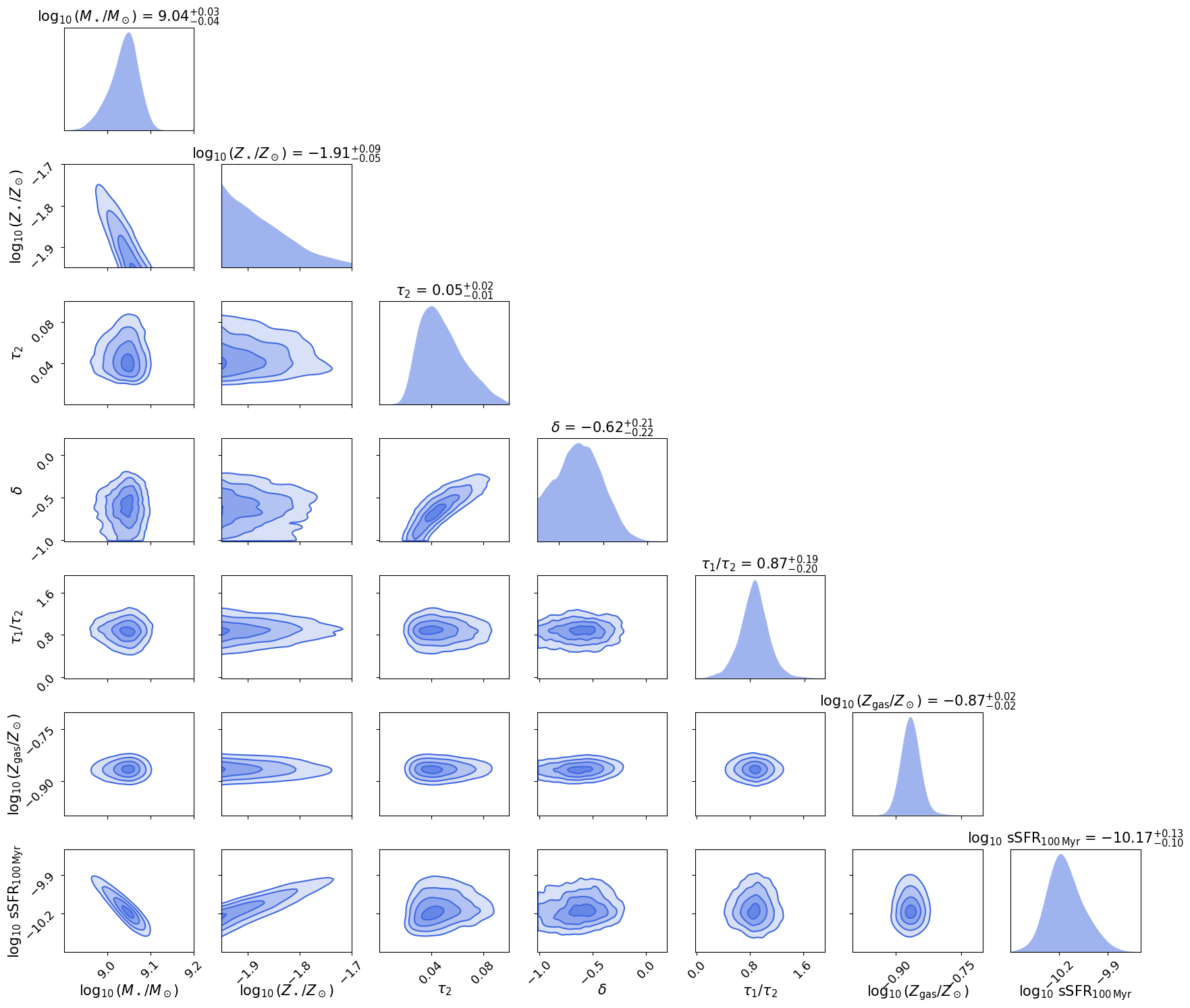}
    \caption{Corner plot showing the marginalized and joint posterior distributions of key host-galaxy parameters derived from the Prospector SED fitting. Parameters include surviving stellar mass ($M_{\ast}$), stellar metallicity ($Z_{\ast}$), dust attenuation parameters ($\tau_2$, the optical depth of diffuse dust; $\delta$, the power-law index of the dust attenuation law; and $\tau_1/\tau_2$, the optical depth of the stellar birth-cloud dust as a fraction of the diffuse dust optical depth), gas-phase metallicity $Z_{\rm gas}$, and the sSFR averaged over the past 100 Myr. The red point marks the best-fit solution.}
    \label{fig:corner}
\end{figure*}

The posterior distributions of the host-galaxy parameters inferred from the {\tt Prospector} fit are shown in Figure~\ref{fig:corner}. The fit yields a low-mass, metal-poor host galaxy, with a surviving stellar mass of $\log(M_\star/M_\odot) = 9.04^{+0.03}_{-0.04}$.  The inferred dust attenuation is low, with $A_V = 0.09^{+0.03}_{-0.02}$, and close to the central extinction value estimated from the Balmer decrement ($A_V = 0.06^{+1.02}_{-0.74}$ assuming $R_V = 3.1$). The specific star formation rate (sSFR) averaged over the past 100~Myr, $ \log_{10}(\rm sSFR/yr^{-1})= -10.17^{+0.13}_{-0.10}$, is consistent with a star-forming galaxy at this redshift.  The low stellar metallicity of $\log(Z_\star/Z_\odot) = -1.91^{+0.09}_{-0.05}$ includes plausible fits that extend to our minimum allowed value of $\log(Z_\star/Z_\odot) = -2$.

\begin{figure*}
    \centering
    \includegraphics[width=0.75\linewidth]{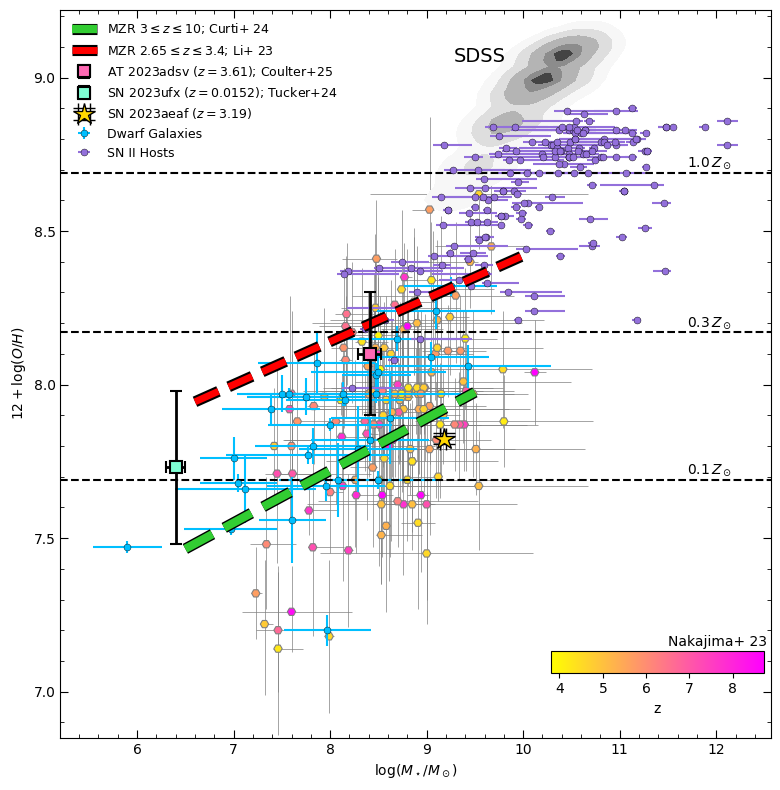}
    \caption{Gas phase metallicity versus stellar mass for the host galaxy of \snname\ (yellow star) compared to literature samples. The pink square star denotes the SN\,II host from \cite{coulter2025discovery} at \textit{z}=3.61, and the cyan square denotes SN 2023ufx, a low-metallicity SN\,II host \citep{Tucker2024_2023ufx}. 
    Colored points show dwarf galaxies (blue; \citealp{Berg12}), low-$z$ SN\,II host galaxies (purple; \citealp{Kelly12}), and {\it JWST}-selected galaxies from $3 < z < 9$ \citep[yellow to pink;][]{Nakajima2023}. The green and red dashed lines show mass–metallicity relations (MZR) at $3\leq z \leq 10$ \citep{Curti2024} and $2.65\leq z\leq 3.4$ \citep{Li2023}, respectively. Horizontal dashed lines indicate constant metallicity levels relative to Solar.  The gray contours show the $z < 0.7$ SDSS galaxy population \citep{Aihara11,Eisenstein11}. Based on Figure 5 from \citet{coulter2025discovery}.}
    \label{fig:coulter_fig}
\end{figure*}

  We convert the {\tt Prospector}-derived gas-phase metallicity, $\log_{10}(Z_{\rm gas}/Z_\odot) = -0.87 \pm 0.02$, to the commonly used $12+\log(\mathrm{O/H})$ scale by adopting the solar abundance from \citet{Asplund2009}, allowing us to place the host of \snname\ on the stellar mass--metallicity relation (MZR) at high redshift as shown in Figure~\ref{fig:coulter_fig}. This yields $12+\log(\mathrm{O/H}) = 7.82^{+0.02}_{-0.02}$. For comparison, the independent $R_{23}$ diagnostic measured from the emission-line fluxes in Section~\ref{host_spectral_analysis} give $12+\log(\mathrm{O/H}) = 8.07^{+0.29}_{-0.20}$ using the calibration by \citep{Mcgaugh1991} and $12+\log(\mathrm{O/H}) = 7.535^{+0.293}_{-0.203}$ using the calibration byb\cite{hirschmann2023}. The estimates are consistent within their uncertainties and indicate a subsolar, relatively unevolved chemical environment for the host galaxy.

To assess the impact of including spectroscopic information on the inferred host-galaxy properties, we compare these results to those obtained from the photometry-only {\tt Prospector} fit. We find that  the stellar mass, dust attenuation, and stellar metallicity are broadly consistent within uncertainties between the two fits, indicating that these quantities are primarily constrained by the broadband photometry. The surviving stellar mass is slightly lower,  $\log(M_\star/M_\odot) = 8.90^{-0.11}_{+0.08}$ in the photometry-only fit, but remains consistent at the $\sim1.5\sigma$ level. In contrast, the gas-phase metallicity shifts to slightly higher values in the absence of spectroscopic constraints, reflecting the sensitivity of nebular emission lines to gas-phase abundances.

 In Figure \ref{fig:coulter_fig}, we  include a $z=3.61$ SN from \cite{coulter2025discovery} as a comparison point for a SN at a similar redshift as well as a low-metallicity, low-$z$ SN\,II at $z \sim 0.1Z_\odot$ from \cite{Tucker2024_2023ufx}.  Notably, the inclusion of the {\it JWST} NIRSpec data significantly improve the errors on gas-phase metallicity compared to the previous SN hosts.
The location of \snname's host is broadly consistent with the extrapolation of the high-redshift MZR \citep[e.g.,][]{Li2023, Curti2024} and overlaps with the population of low-mass, star-forming dwarf galaxies. Compared to samples of SN~II host galaxies at low redshift, the host of \snname\ lies at systematically lower metallicity, reflecting the unevolved chemical environment of galaxies at $z \sim 3$.

\section{Progenitor Modeling} \label{sec:progenitor_modeling}

We next model the SN progenitor by adopting the same methodology used in recent {\it JWST} studies of high-redshift SNe\,II \citep{coulter2025discovery,moriya2025}.  We generate model SEDs using the one-dimensional, multi-frequency radiation-hydrodynamics code \texttt{STELLA} \citep{Blinnikov1998, Blinnikov2000, Blinnikov2006}, and integrate these through the {\it JWST} filter transmission curves in the observer frame. The models vary explosion parameters that include the Zero-Age Main Sequence (ZAMS) mass, the explosion energy, and the synthesized $^{56}$Ni mass, and are evaluated both with and without the presence of a confined dense CSM. Extinction from the host galaxy is included as a free parameter in the fits and constrained to be consistent with the Balmer decrement measurement; this provides a looser constraint than the {\tt Prospector} results, and allows for modest differences between the nebular extinction compared to the local ISM.

\begin{figure}
    \centering
    \includegraphics[width=1.0\linewidth]{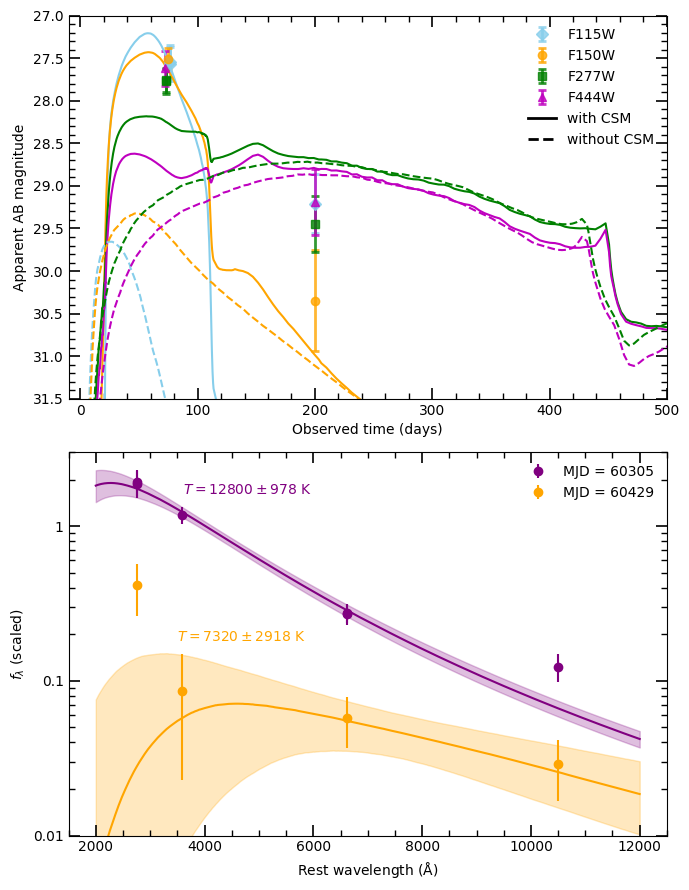}
    \caption{Top panel: Observed {\it JWST} light curves of \snname\ compared to SN\,II models with (solid) and without (dashed) CSM interaction.  Bottom panel: Rest-frame SEDs at the two observed epochs. The colored curves show the median blackbody fits derived from the posterior distributions, while the shaded regions indicate the corresponding $1\sigma$ uncertainties. 
    The second-epoch {\it F115W} flux lies above the simple blackbody prediction, possibly indicating deviations from a pure thermal continuum at shorter wavelengths or an extended CSM density profile.}
    \label{fig:model_plot}
\end{figure}

In Figure \ref{fig:model_plot} (bottom panel) we show that the first epoch is consistent with a high effective temperature ($\gtrsim 10{,}000$~K), which can be explained by early interaction between the SN ejecta and a dense, confined CSM; observed CSM interaction appears to be relatively common in the high-redshift CC\,SNe discovered to date \citep[e.g.,][]{coulter2025discovery,moriya2025}, and may be especially prevalent observationally due to time dilation and the redshifting of rest-frame UV emission into the observed bands. In contrast, the second epoch is broadly consistent with a cooler temperature of $\sim 7000$~K, near the hydrogen recombination temperature expected during the plateau phase of SNe\,II. These results suggest an evolutionary transition from an interaction-dominated phase to a recombination-powered plateau.

Light-curve modeling supports this interpretation, with the first epoch consistent with a dense CSM confined within $\sim 10^{15}$~cm, corresponding to a mass-loss rate of $\sim 10^{-2}\,M_\odot\,\mathrm{yr^{-1}}$ for a $10~\mathrm{km\,s^{-1}}$ wind, and a total CSM mass of $\sim 0.5\,M_\odot$. The second epoch is consistent with the plateau phase of a canonical SN~II resulting from a $\sim 10^{51}$~erg explosion of a $\sim 12\,M_\odot$ progenitor. However, with only two epochs of observation, it is not possible to place strong constraints on the progenitor mass, with plausible fits extending up to $\sim24~M_{\odot}$, or detailed explosion properties.

These inferred values are broadly consistent with those reported by \cite{moriya2025}, who find that mass-loss rates of $\sim10^{-3}$–$10^{-2}\,M_\odot,\mathrm{yr^{-1}}$ and compact CSM confined within $\sim10^{15}$–$10^{16}$ cm are typical of red supergiant progenitors undergoing enhanced pre-SN mass loss. Our inferred mass-loss rate of $\sim10^{-2}\,M_\odot,\mathrm{yr^{-1}}$  therefore places this event toward the upper end of the expected range.

An excess flux that is not explained by the model is also observed in the {\it F115W} band at the second epoch. While this feature could reflect residual CSM interaction, particularly if the CSM density profile transitions smoothly rather than exhibiting a sharp outer boundary, alternative explanations such as modest deviations from a simple blackbody or temperature gradients within the ejecta cannot be ruled out. Importantly, comparisons with nearby SNe\,II (Figure \ref{fig:swift_comp}) indicate that the ultraviolet color evolution is not anomalous at this phase.

\section{Discussion and Conclusions} \label{discussions and conclusions}
In this study, we report the discovery and photometric classification of \snname, a $z = 3.19$ transient detected in the COSMOS-Web survey. \snname\ has one of the highest  spectroscopic redshifts of any SN discovered and characterized to date, surpassed by only by a small number of other {\it JWST}-discovered SNe. Light curve-based classification using \textit{JWST} photometry, together with comparisons to nearby SNe observed by \textit{Swift}, indicates that the photometric evolution of \snname\ is most consistent with a SN~II origin.

We analyze the host-galaxy spectrum and photometry of \snname, which provides important context for its environment. Its Balmer decrement measurement indicates no significant internal dust attenuation assuming case B recombination. The $R_{23}$ diagnostic suggests a chemically unevolved, low-metallicity interstellar medium at the SN location.
Joint photometric and spectroscopic SED fitting with {\tt Prospector}, as seen in Figure \ref{fig:model_plot}, confirms this result, finding that the host of \snname\ is a low-mass, metal-poor, star-forming host galaxy, with a surviving stellar mass of $\log(M_\star/M_\odot) = 9.04^{+0.03}_{-0.04}$, a sSFR of $ \log_{10}(\rm sSFR/yr^{-1})= -10.17^{+0.13}_{-0.10}$, a low stellar metallicity consistent with $\log(Z_\star/Z_\odot) \simeq -2$, and a gas-phase metallicity of $\log(Z_{\text{gas}}/Z_\odot) = -0.87\pm0.02$. The host lies along the expected $z = 3$ MZR and overlaps with low-mass, star-forming dwarf galaxies (albeit somewhat on the more massive side of the dwarf galaxy distribution). Overall, the host properties are typical of chemically young, low-metallicity environments where massive-star explosions are expected at early cosmic times.

We next model the progenitor of \snname\ using \texttt{STELLA}. 
Light-curve modeling favors a confined CSM within $\sim10^{15}$~cm, corresponding to a mass-loss rate of $\sim10^{-2}\,M_\odot\,\mathrm{yr^{-1}}$ and a total CSM mass of $\sim0.5\,M_\odot$. The second epoch is consistent with the plateau phase of a $\sim10^{51}$~erg explosion of a $\sim12\,M_\odot$ progenitor, but with plausible fits extending as high as 24\,$M_{\odot}$. The high confined CSM mass inferred for the host may be consistent with expectations for massive stars in low-metallicity environments, where reduced line-driven winds can allow progenitors to retain more of their hydrogen envelopes before explosion. However, the limited two-epoch coverage prevents strong constraints on the progenitor mass, metallicity-dependent mass loss, or detailed explosion properties.

Cumulatively, \snname\ exhibits several features that appear to be increasingly common at early cosmic times. The very blue, hot emission at the first epoch is consistent with other high-$z$ candidates that show evidence for strong early-time  circumstellar interaction (e.g., 
\citealp{moriya2025}). The low-mass, metal-poor host environment of \snname\ is also consistent with expectations for massive-star explosions in chemically young galaxies at $z \sim 3$. Lastly, the luminosity of \snname\ is higher than typical for local-Universe SNe\,II, which may be characteristic of CC~SNe with low-metallicity progenitors \citep{Anderson18,Gutierrez20,Tucker2024_2023ufx}.  However, the limited temporal coverage restricts detailed constraints on progenitor properties, highlighting the need for larger samples to robustly characterize the high-redshift SN\,II population.

This work adds another SN to the growing set of CC\,SNe observed at $z > 3$, and demonstrates the power of \textit{JWST} for identifying and characterizing such SNe at high redshift. 
\snname's redshift places it near the peak of cosmic star formation, where the large population of massive and short-lived stars provide the potential for further CC\,SN discoveries in the coming years.  As the sample grows, the rates of such SNe can be used to place independent constraints on the cosmic star-formation history, the variability of the high-mass end of the stellar initial mass function, and the evolution of CC\,SN properties with redshift.
 
Forthcoming observations with the {\it Nancy Grace Roman Space Telescope} will extend this work by discovering large samples of high-redshift CC\,SNe, opening the door to statistical studies of their properties and environments \citep{rose2021referencesurveysupernovacosmology}.  However, $z \gtrsim 3$ SNe\,II like \snname\ will likely be too faint for {\it Roman}, making every {\it JWST} discovery critical for understanding this population.

\begin{acknowledgments}
D.O.J.\ and V.A.\ acknowledge support from NSF grants AST-2407632, AST-2429450, and AST-2510993, NASA grants 80NSSC24M0023 and 80NSSC24K0353, and HST/JWST grants HST-GO-17128.028 and JWST-GO-05324.031, awarded by the Space Telescope Science Institute (STScI), which is operated by the Association of Universities for Research in Astronomy, Inc., for NASA, under contract NAS5-26555.  This work is also funded in part by the Gordon and Betty Moore Foundation through Grant GBMF13900 to D.O.J.

W.B.H. acknowledges support from the National Science Foundation Graduate Research Fellowship Program under Grant No. 2236415.

B.W. acknowledges support provided by NASA through Hubble Fellowship grant HST-HF2-51592.001 awarded by the Space Telescope Science Institute, which is operated by the Association of Universities for Research in Astronomy, In., for NASA, under the contract NAS 5-26555. The STScI TSST group acknowledges partial support
from JWST-GO-06541, JWST-GO-06585, and JWST-
GO-05324.

 N.E.D acknowledges support from NSF grants LEAPS-2532703 and AST-2510993.

\end{acknowledgments}

\facility{{\it JWST} (NIRCam, NIRSpec) }

\software{{\tt Prospector} \citep{leja_prospector, johnson_prospector}, {\tt STARDUST2} \citep{rodney_stardust2}, {\tt emcee} \citep{Foreman_Mackey_2013}, FSPS \citep{conroy_2009}, {\tt sncosmo} \citep{barbary2014sncosmo}.}

\bibliography{sample701}{}
\bibliographystyle{aasjournalv7}

\end{document}